# Microbial Genome as a Fluctuating System: Distribution and Correlation of Coding Sequence Lengths

**Vasile V. Morariu**

Department of Molecular and Biomolecular Physics
National Institute of Research and Development for Isotopic and Molecular Technology,
400293 Cluj-Napoca , Romania e-mail: vvm@itim-cj.ro

**Abstract** The length of coding sequence (CDS) series in microbial genomes were regarded as a fluctuating system and characterized by the methods of statistical physics. The distribution and the correlation properties of 50 genomes including bacteria and archaea were investigated. The distribution of lengths was investigated by rank size analysis (Zipf's law) as the DNA is commonly regarded as a language. However we found that CDS series does not obey Zipf's law as for natural languages. The distribution was found to be more closely to an exponential distribution although in some cases such a distribution was only a first degree approximation. The correlation organization was investigated by detrended fluctuation analysis (DFA). The DFA plot revealed two different correlation exponents at short distance between CDS and at more distant intervals respectively. Generally the short distance correlation was slightly above 0.5 which indicates a residual correlation. The correlation at more distant terms was either higher or lower than 0.5 depending on the species. Some of the genome data was similar to DFA characteristics for English and Greek and for Lynux computer language. Segmentation of the whole genome series into ten shorter series proved that the variability of distribution, for these segments, is significantly diverse and confirm the non uniform organization of the length series. This is also revealed by the correlation analysis. The higher correlation corresponds to segments characterized by higher variability.

## 1. Introduction

The microbial genomes consist mainly of coding sequences (CDS) while non-coding sequences represent a minor proportion of the genome. Consequently the composition of coding genes, comprise most of the genome (~90%) and the bacterial genome size shows a strong positive relationship with gene number. The length of a CDS ranges between several hundreds and several thousands of base pairs. A typical species of bacteria, like *Escherichia coli* contains around $5x10^6$ base pairs which are organized in several thousands of genes. The series of CDS lengths can be regarded as a space fluctuating series and therefore are suitable for the statistical analysis. It should be mentioned that while the correlation properties of DNA at the level of bases have been intensively studied for more than a decade, the series of coding sequence (CDS) lengths received very little attention. Although correlation characterization of CDS length series has been previously attempted on bacterial genomes, a full statistical characterization has not been reported to date [1-3]. While the older papers on the subject suggested a weak, long–range correlation the more recent papers, at contrary, brought evidence for short-range correlation and generally a non-uniform organization [3-4]. A systematic investigation of the CDS length distribution remained an open issue.

On a parallel line, a genome sequence is a one-dimensional letter string composed of four types of letters (bases), A, T, G and C, and thus has characteristics appropriate for being called "language." The elucidation of language that rules genome was regarded as a new scientific field [5].

The aim of this paper was to establish distribution and correlation characteristics of CDS length series in the genome of selected species and strains of bacteria and archaea. The investigation of distribution relies upon the rank-size distribution analysis and the related Zipf's law. This implies a linguistic like approach to the CDS length series. The main tool for the correlation investigation was detrended fluctuation analysis (DFA) and segmentation of the genome series of data.

## 2. Materials and Methods

A total number of 50 genomes of bacteria and archaea were investigated (Table 1). The species were selected such as to cover all main divisions of microbes as well as some of the main models used in literature (*E. coli* and *B. subtilis*). In case of *E. coli* and *Staphilococus aureus* various strains were included in order to compare the variability of the statistical properties of strains.

The extraction of the data, from the EMBL-EBI data base, was done with a program written in MATLAB. The length of coding sequences, expressed as number of base pairs, was calculated as the difference between the start and the end position of CDS in the genome. The mean value of the CDS length is also available in the proteome section of EMBL-EBI data base. Variability of the CDS length series can be estimated by calculating the standard deviation of the series.

The distribution of the CDS lengths was analyzed by probability density of lengths, and rank size analysis. A probability density investigation presents the data as frequency counts *versus* bin centre while the width of the bin is either automatically or manually selected. The procedure is performed by the ORIGIN program. Rank-size distribution or the rank-size rule (or law) describes the regularity in many phenomena including the distribution of city sizes around the world, sizes of businesses, particle sizes, lengths of rivers, frequencies of word usage, wealth among individuals, etc. [6]. All are real-world observations that follow power laws. If one ranks the word usages for a given language and calculates the natural logarithm of the rank and of the word usage, the resulting graph will show a remarkable log-linear pattern. This is the rank-size distribution. Therefore the law (or distribution) takes the form:

$$P(r) \propto r^{-\alpha} \qquad (1)$$

where *P(r)* is the normalized frequency of a word whose rank is *r*. When such a distribution refers to natural languages the slope is close to one (α=1) and it is commonly known as Zipf's law. We found that our CDS length series however obey an exponential law. A continuous random variable X is said to have an exponential distribution if it has probability density function:

$$f_X(x,\lambda) = \lambda e^{-\lambda x} \text{ for } x \rhd 0 \qquad (2)$$
$$0 \quad \text{for } x \le 0$$

where λ > 0 is called the rate of the distribution. It can be seen that as λ gets larger, the process to happen tends to happen more quickly, hence we think of λ as a rate.

The correlation properties of the series were investigated by detrended fluctuation analysis (DFA). The DFA method was originally developed to investigate long-range correlation in non-stationary series [7]. First DFA integrates the series which is further divided into boxes of equal length, *n*. A least square line is fit to the data in each box *n* which represents the trend in that box. The integrated time series *y(k)* is detrended by subtracting the local trend $y_n(k)$ in each box. Then the root-mean square of the resulting series is calculated as a fluctuation function:

$$F(n) = \sqrt{\frac{1}{N}\sum_{1}^{N}[y(k) - y_n(k)]^2} \qquad (3)$$

Here *N* is the number of terms. *F(n)* typically increases with box size *n* and a linear relationship on a double log graph indicates the presence of scaling:

$$F(n) \propto n^{\alpha} \qquad (4)$$

The outcome of DFA analysis is the correlation exponent α. A single correlation exponent over at least two orders of magnitude of box sizes *n* describes long range correlation properties, while multiple correlation exponents generally describe short-range properties of the fluctuating system [8]. Their values range between $0.5<\alpha<1.5$ and $\alpha<0.5$ for anti correlated cases. When $\alpha=0.5$ the terms of the series are uncorrelated. The key step in a long-range correlation decision is that DFA plot should be linear over the whole range of box sizes covering the series. However the DFA plot of CDS length series was generally non-linear and it was approximated by two straight lines described by the slopes $\alpha_1$ and $\alpha_2$. The DFA results were checked against the shuffled series.

The results for the distribution and correlation analysis are included in table 1 so that individual results can be located in figures 3, 4 and 6.

### 3. The Distribution Characteristics

The classical example of Zipf's law application are the natural languages where the exponent $\alpha=1$ [9]. It has been argued that random text models also obey Zipf's law and therefore Zipf's law is meaningless for linguistic analysis. A careful analysis however revealed that meaningfulness of Zipf's law is high [9]. Although the differences between genome and language are quite important it has been considered useful to make an analogy between language evolution and genome evolution [10].

Obviously different words having the same length may exist in a natural language yet the difference among them is given by their meaning. In our case, a "word" is assimilated to a CDS and its particular identity is given by its length. The frequency of individual words of natural languages is strictly speaking a different matter compared to a CDS. A CDS is unique and it has practically a single appearance in the genome, while a word can be repeated many times in text. As the frequency count of CDS length refers to intervals of values, many different but closely similar lengths are "melted" in a single frequency. This means that an analogy with the natural languages may result when to a certain range of lengths are attached a single frequency. Consequently a Zipf's analysis applied to CDS length series is an approximation of the natural linguistic case. The basic difference is that the frequency of a word is replaced by the frequency count of an interval of CDS lengths. This makes possible a comparison between a natural language and a genomic language.

Figure 1 illustrates Zipf's plot for two bacterial genomes. These plots do not obey a power law. The Zipf's plot for a series of normal random distributed numbers is illustrated on the same plot. The main difference between the genome data and the random numbers is the longer plateau for the latter series. The genome series appear more related to a random uncorrelated series and definitely not related to the Zipf's plot of natural languages (which is a single line with $\alpha$=1). Therefore the distribution of "words" in a microbial genome language is qualitatively different from a natural language.

We have also investigated other types of distribution and the closest we found was an exponential like distribution. This was true for some microbial species (figure 2a) while for others such a distribution was only a first degree approximation (figure 2b). Anticipating the results described in sections below, the reason for deviation from an exponential distribution is the fact that the CDS length series of data are non uniform, or put it in another way they are "composite" structures. This aspect will be further exemplified in section 5.

A characteristic parameter of the exponential distribution is the rate constant $\lambda$ which tells how fast the terms of the series change. Its relationship to the variability of the CDS length series for various genomes is illustrated in figure 3. Although these parameters are intuitively related the outcome of this correlation is unexpected.

The plot shows that as variability of the CDS length series for various genomes increases, the rate constant of the exponential distribution decreases. This relationship separates the genomes into at least three or more groups (figure 3). The same variability of a genome may be characterized by a variety of $\lambda$ rate constants. A possible cause for this aspect is that $\lambda$ refers to distributions which are more or less closer to a perfect exponential distribution. A qualitative evaluation of the data in figure 3 seems to suggest that distribution corresponding to higher values of SD correspond to more pronounced deviations from an exponential distribution.

Our entire experience with the CDS length series led us to a non uniform structure hypothesis. It seems likely that the genome data we analyzed is a superposition of different characteristics. As a further characterization of this hypothesis we checked for the variability of CDS length series *versus* the mean CDS length. Variability of the CDS length series can be estimated by calculating the standard deviation of the series. As the mean value of CDS length increases, variability of the series increases too (figure 4). The interesting aspect of this plot is twofold: a) It shows there is a lower and an upper limit for the variability of CDS lengths. These limits seem quite linear (at least the lower limit) and they tell that microbial species below and above these limits cannot exist; b) As the limits delimitate an angular space the lower the mean value of the CDS length the lower the variability which may exist for similar mean values. This points out to a hypothetic organism which has the lowest possible mean value and variability at around x=5500 bp and y=250 bp. The whole picture looks like an evolutionary image by starting from a single most simple species which evolves into more and more diversified species. It is quite possible that the 48 species analyzed in this work cannot give a precise image about the lower and the upper limits or the exact characteristics of the "primordial" species, so that a better delimitation of the angular space awaits more data [11].

Figure 1 Zipf's plot for two different species of bacteria. Zipf's plot for normal random numbers is included for comparison. The size of each series is the similar, 2500 data.

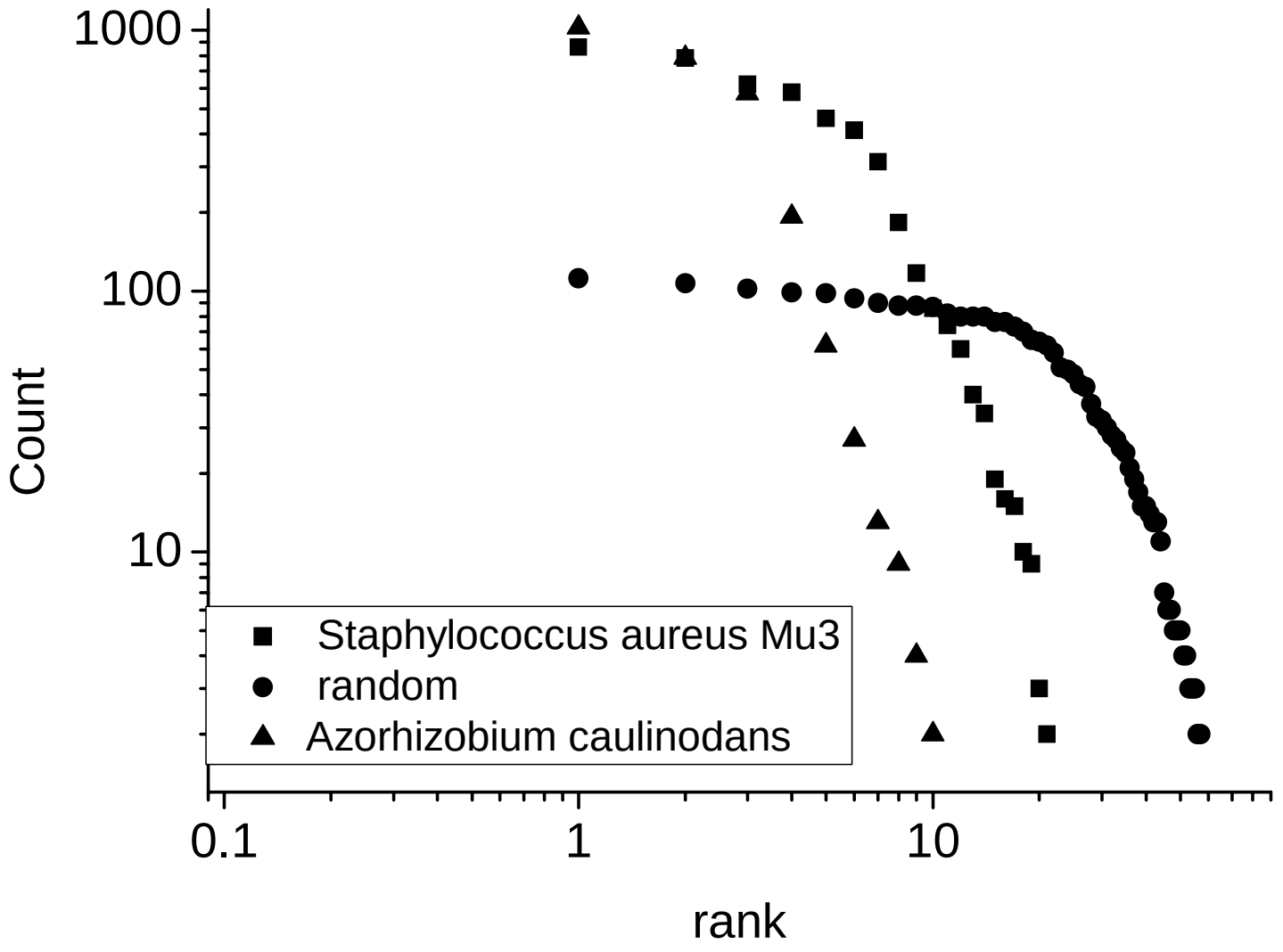

Figure 2 Rank size distribution of CDS length series for two bacterial species. While the upper plot is well described by an exponential distribution, the lower plot is clearly different. The straight line is the exponential approximation. The value of the rate constant λ (the slope of the semi logarithmic plot) is illustrated in the figures.

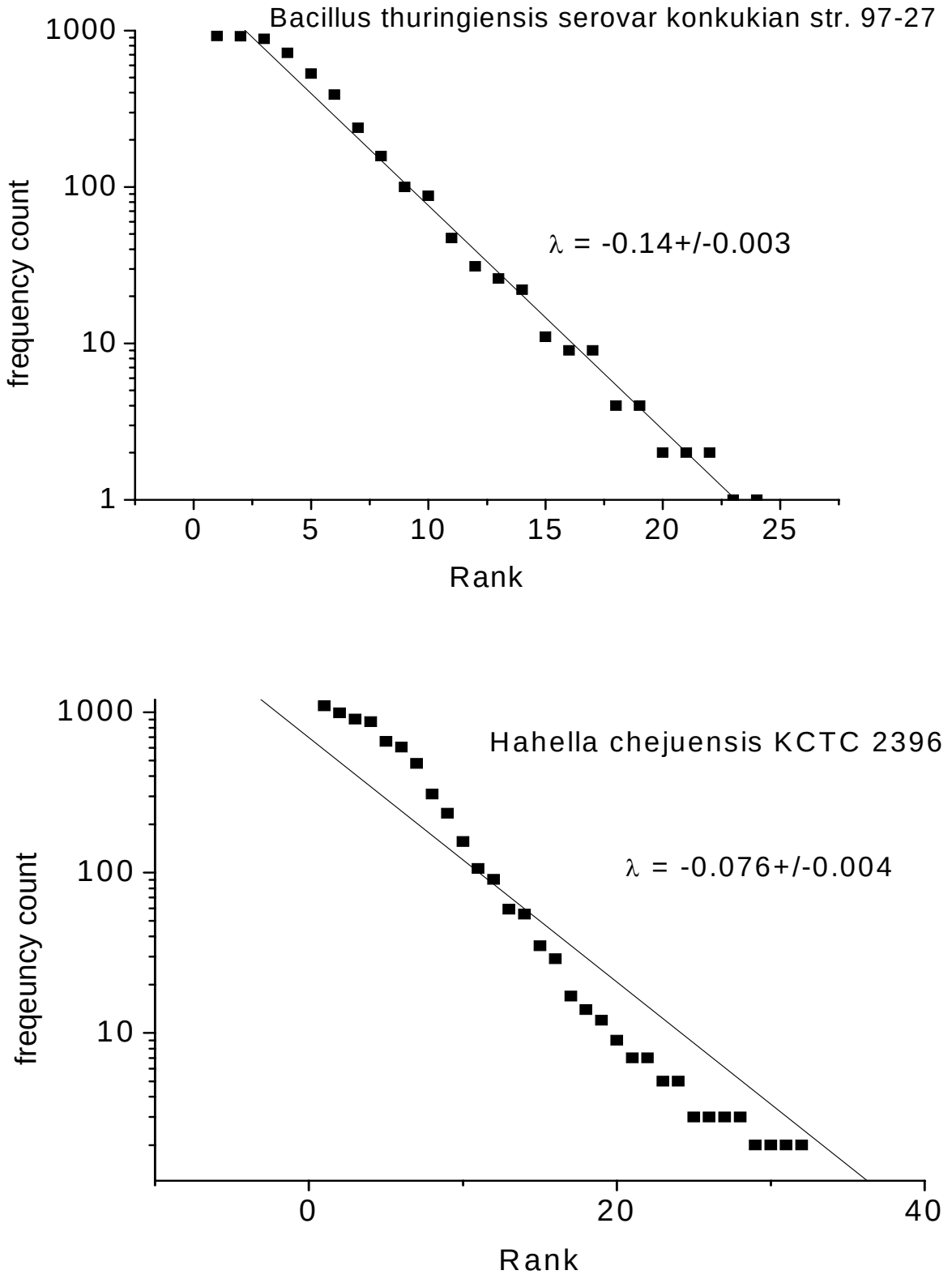

Figure 3. The rate constant λ of the exponential distribution *versus* the variability (SD) of CDS length series for various microbial genomes. The straight lines are intended to guide the eye.

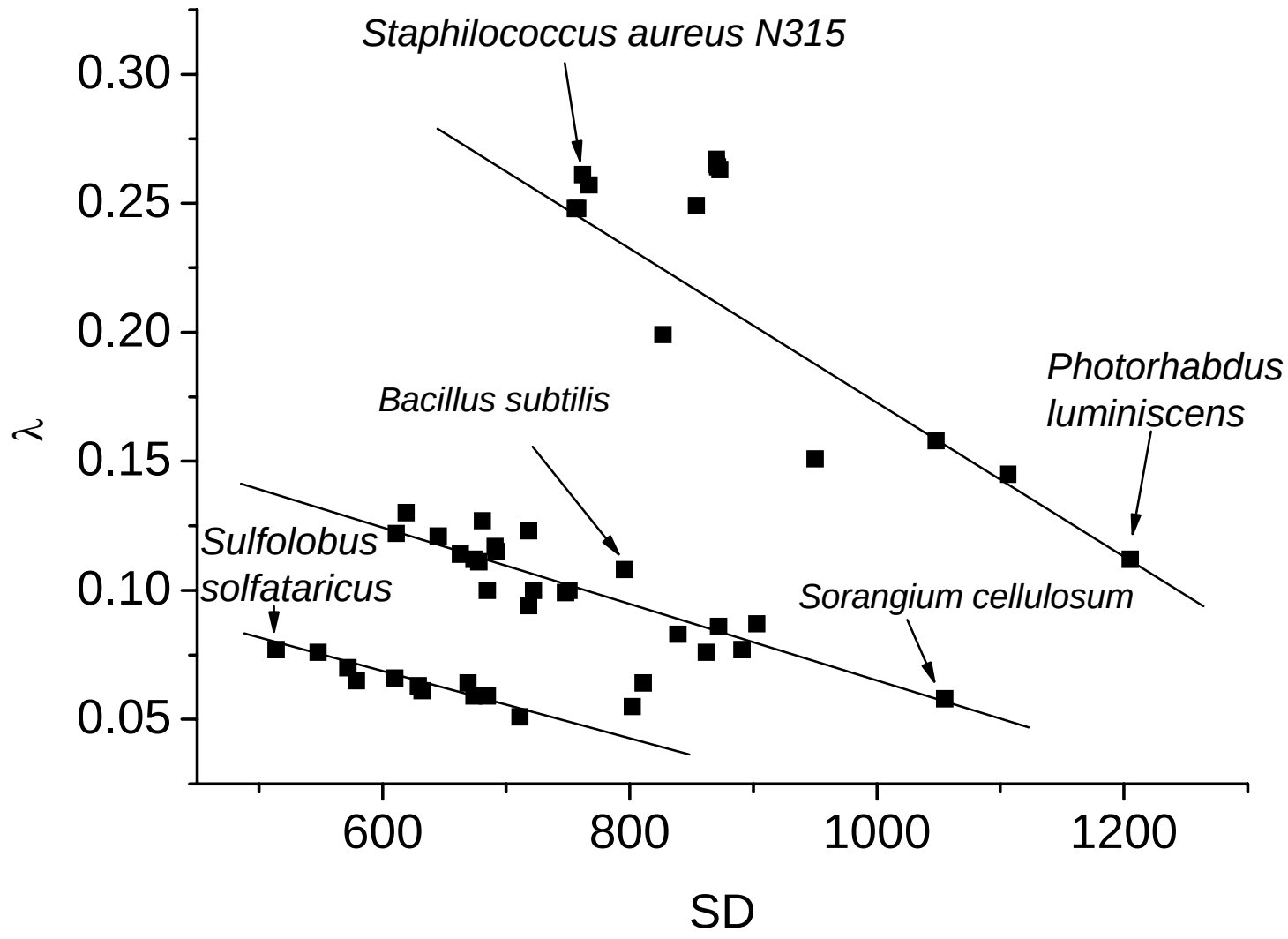

Figure 4 Variability (SD) of the CDS length series *versus* the mean CDS length for 48 microbial genomes. The straight lines guide the eye for the limits of the space covered by the data.

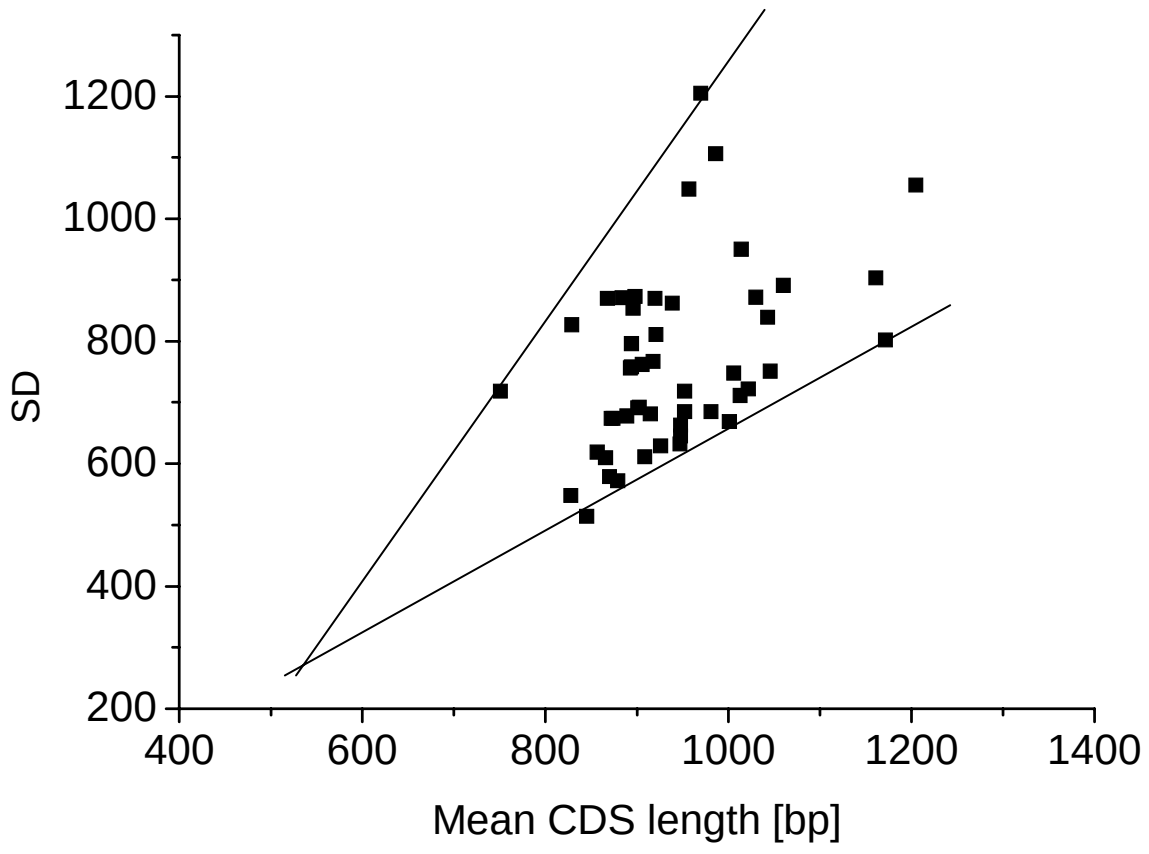

Table 1 Correlation exponent $\alpha_1$ and $\alpha_2$, rate constant $\lambda$ of the exponential distribution and the mean values of the CDS length series for different microbes and several languages.

| **No.** | **Organism/Language** | **Alpha 1** | **Alpha 2** | **λ** | **Mean value± SD** |
|---|---|---|---|---|---|
| 1 | Acidovorax avenae subsp. citrulli AAC00-1 | 0.564±0.001 | 0.564±0.001 | 0.051±0.001 | 1013±711 |
| 2 | Anabaena variabilis ATCC 29413 | 0.538±0.002 | 0.689±0.006 | 0.086±0.002 | 1030± 872 |
| 3 | (Archaea): Aeropyrum pernix K1 DNA | 0.578±0.003 | 0.454±0.005 | 0.065±0.002 | 870±579 |
| 4 | Azorhizobium caulinodans ORS 571 DNA | 0.514±.0.003 | 0.659±0.007 | 0.100±0.005 | 1022±722 |
| 5 | (Archaea): Archaeoglobus fulgidus DSM 4304 | 0.501±0.003 | 0.762±0.018 | 0.076±0.002 | 828±548 |
| 6 | Bradyrhizobium japonicum USDA 110 DNA | 0.527±0.001 | 0.971±0.013 | 0.100±0.005 | 952±685 |
| 7 | Bacteroides thetaiotaomicron VPI-5482 | 0.611±0.006 | 0.676±0.003 | 0.055±0.001 | 1172±802 |
| 8 | Bacillus subtilis | 0.757±0.008 | 0.653±0.014 | 0.108±0.005 | 894±796 |
| 9 | Bacillus pumilus SAFR-032 | 0.644±0.004 | 0.706±0.011 | 0.112±0.005 | 874±674 |
| 10 | Bacillus halodurans C-125 DNA | 0.561±0.002 | 0.710±0.012 | 0.070±0.001 | 879±572 |
| 11 | Bacillus thuringiensis serovar konkukian str. 97-27 | 0.511±0.001 | 0.753±0.013 | 0.130±0.004 | 857±619 |
| 12 | Clostridium beijerinckii NCIMB 8052 | 0.525±0.002 | 0.641±0.004 | 0.121±0.004 | 948±645 |
| 13 | Delftia acidovorans SPH-1 | 0.555±0.004 | 0.633±0.002 | 0.099±0.003 | 1006±748 |
| 14 | Escherichia coli O157:H7 Sakai | 0.558±0.002 | 0.757±0.009 | 0.117±0.002 | 901±691 |
| 15 | Escherichia coli APEC O1 | 0.557±0.001 | 0.669±0.014 | 0.064±0.001 | 1001±669 |
| 16 | Escherichia coli O157:H7 EDL933 | 0.553±0.003 | 0.738±0.005 | 0.115±0.003 | 903±692 |
| 17 | Escherichia coli CFT073 | 0.551±0.003 | 0.720±0.010 | 0.059±0.001 | 877±674 |
| 18 | Escherichia coli K12 | 0.541±0.026 | 0.599±0.004 | 0.061±0.001 | 947±632 |
| 19 | Enterococcus faecalis V583 | 0.542±0.002 | 0.599±0.005 | 0.111±0.003 | 889±678 |
| 20 | English | 0.480± | 0.650± | - | - |
| 21 | Flavobacterium johnsoniae UW101 | 0.585±0.006 | 0.636±0.004 | 0.073±0.003 | 1060±891 |
| 22 | Frankia sp. EAN1pec | 0.589±0.002 | 0.520±0.006 | 0.083±0.003 | 1043±839 |
| 23 | Greek | 0.460± | 0.610± | - | - |
| 24 | Hahella chejuensis KCTC 2396 | 0.573±0.001 | 0.543±0.005 | 0.076±0.004 | 939±862 |
| 25 | Haemophilus influenzae (strain 86-028NP) | 0.525±0.002 | 0.712±0.005 | 0.063±0.002 | 926±629 |
| 26 | Haemophilus influenzae (strain ATCC 51907 / KW20 / Rd) | 0.538±0.005 | 0.507±0.004 | 0.064±0.001 | 921±811 |
| 27 | Haemophilus influenzae Pitt EE | 0.562±0.008 | 0.649±0.003 | 0.264±0.017 | 884±871 |
| 28 | Helicobacter pylori | 0.504±0.004 | 0.553±0.003 | 0.123±0.004 | 952±718 |
| 29 | Lactobacillus plantarum strain WCFS1 | 0.516±0.002 | 0.639±0.008 | | |
| 30 | Lactobacillus casei ATCC 334 | 0.538±0.001 | 0.58±00.004 | 0.066±0.002 | 866±610 |
| 31 | Linux | 0.640 | 0.850 | - | - |
| 32 | Mycoplasma penetrans HF-2 DNA | 0.661±0.002 | 0.416±0.016 | 0.087±0.002 | 1161±903 |
| 33 | Mycoplasma pneumoniae M129 | 0.542±0.003 | 0.601±0.006 | 0.100±0.003 | 1046±751 |
| 34 | Mycobacterium smegmatis str. MC2 155 | 0.523±0.005 | 0.502±0.002 | 0.114±0.004 | 948±663 |
| 35 | Microcystis aeruginosa NIES | 0.568±0.004 | 0.643±0.005 | 0.094±0.004 | 751±718 |
| 36 | (Archaea) : Methanococcoides burtonii DSM 6242 | 0.517±0.001 | 0.544±0.013 | 0.122±0.004 | 909±611 |
| 37 | Nocardia farcinica IFM 10152 DNA | 0.560±0.003 | 0.471±0.008 | 0.158±0.016 | 957±1048 |
| 38 | Pseudomonas entomophila str. L48 | 0.572±0.003 | 0.504±0.001 | 0.151±0.013 | 1014±950 |
| 39 | Photorhabdus luminescens subsp. laumondii TTO1 | 0.584±0.002 | 0.632±0.007 | 0.112±0.011 | 970±1205 |
| 40 | Rhodococcus sp. RHA1 | 0.551±0.001 | 0.464±0.004 | 0.145±0.015 | 986±1106 |
| 41 | Sulfolobus solfataricus | 0.583±0.003 | 0.481±0.003 | 0.077±0.002 | 845±514 |
| 42 | Sorangium cellulosum 'So ce 56' | 0.683±0.002 | 0.556±0.005 | 0.053±0.003 | 1205±1055 |
| 43 | Staphylococcus aureus subsp. aureus JH1 | 0.524±0.001 | 0.639±0.007 | 0.264±0.017 | 884±871 |
| 44 | Staphylococcus aureus subsp. aureus JH9 | 0.528±0.001 | 0.634±0.008 | 0.263±0.017 | 898±873 |
| 45 | Staphylococcus aureus subsp. aureus Mu3 | 0.572±0.003 | 0.577±0.008 | 0.248±0.018 | 894±758 |
| 46 | Staphylococcus aureus subsp. aureus Mu50 | 0.582±0.004 | 0.536±0.003 | 0.248±0.019 | 893±756 |
| 47 | Staphylococcus aureus subsp. aureus MW2 | 0.529±0.001 | 0.595±0.010 | 0.249±0.018 | 896±854 |
| 48 | Staphylococcus aureus subsp. aureus N315 | 0.578±0.004 | 0.577±0.005 | 0.261±0.017 | 906±762 |
| 49 | Staphylococcus aureus subsp. aureus NCTC 8325 | 0.526±0.001 | 0.567±0.005 | 0.199±0.013 | 829±827 |
| 50 | Staphylococcus aureus subsp. aureus str. Newman | 0.569±0.003 | 0.589±0.004 | 0.257±0.017 | 918±767 |
| 51 | Staphylococcus aureus subsp. aureus USA300 FPR3757 | 0.525±.0021 | 0.596±0.007 | 0.265±0.017 | 920±870 |
| 52 | Staphylococcus aureus subsp. aureus USA300 TCH1516 | 0.528±0.001 | 0.638±0.013 | 0.267±0.018 | 868±870 |
| 53 | Xanthobacter autotrophicus Py2 | 0.580±0.003 | 0.5600±.003 | 0.059±0.001 | 981±685 |

## 4. The correlation characteristics

The DFA plot for a microbial species and the corresponding shuffled series are illustrated in figure 5. It can be seen that at shorter distances (below about one order of magnitude) the correlation exponent $\alpha_1$ has a value close to 0.5 yet slightly higher which indicate a residual correlation. At higher distances among the terms of the series correlation increases to a significantly higher value $\alpha_2$. The shuffled series remain close to the uncorrelated value of 0.5. Further the whole picture of the correlation exponents is illustrated in figure 6 for all genomes. Included are the values corresponding to two natural languages and a computer language. We found that the general shape of the DFA plot for microbial genomes resembled the similar plots for these languages [12]. At the same time the value of the correlation exponents for shorter and longer distances placed the languages within the area covered by the CDS length series of microbial genomes. This striking similarity was the actual reason for searching possible connection between the natural languages and the microbial languages. However the additional distribution analysis revealed the clear difference between the natural and the microbial languages.

It should be further mentioned that the kind of DFA plots illustrated in figure 5 was not seen for all genomes. Some of the species presented DFA plots with a downwards bend instead of upwards as in figure 5. However none of the investigated microbes showed a linear plot over the whole range of boxes *n*. At the same time the upwards bend appeared as a constant feature for the natural and computer languages.

Although the short distance correlation characteristic as reflected in the value of $\alpha_1$ is close to the random 0.5 value, its value seems significantly different from 0.5. This is further illustrated in figure 6 where $\alpha_2$ is plotted against $\alpha_1$ for different microbial genomes. This correlation "phase space" like picture shows the variability of correlation. Included are the corresponding values for English and Greek languages as well as for Lynux computer language taken from ref. 12. They all seem to generally fit into the area of microbial correlation data. Particular features of the natural languages seem the lowest value of the $\alpha_1$ correlation while Lynux has the highest $\alpha_2$ value. *Bacillus subtilis* and *Mycoplasma penetrans* has extreme values for $\alpha_1$ and $\alpha_2$ respectively. On the other hand the distribution characteristics of these bacteria are placed into the middle of figure 3. Obviously the distribution and correlation characteristics are independent. Putting together the distribution and correlation data we can notice both similarity and difference among the natural and genome languages. The main similarity lays in the correlation property while distribution is qualitative distinct.

## 5. Segmentation analysis

If any random series is segmented into very short intervals may reveal local correlations. These are the natural local fluctuation of the series. However they disappear when looking at longer segments where lack of correlation is dominant. The opposite is a long range correlated series where the same correlation is seen on all the scales of the series. The segments of such a series show much the same long range correlation. Obviously the segments must remain long enough to avoid the correlation fluctuation arising when very short segments are considered. Such very short segments may represent intervals consisting of anything between 10 and 100 terms. The CDS length series on the other hand are comparatively long. For example *Hahella chejunesis* contains 6782 CDS terms and *Bacillus thuringiensis* is a 5117 terms long series. The idea of segmentation analysis is intended to check out whether the organization of the series is non uniform. Some preliminary correlation investigations on segmentation were already published for *Bacillus subtilis* and *Escherichia coli* [3-4]. Non uniform organization could be revealed by segmenting the series into reasonably long series and by looking at their distribution and correlation characteristics. A uniform structure should keep constant the statistical features while a non uniform structure should reveal differences among segments. We have divided the above series into ten segments so they are more 500 terms long. Such series are clearly longer than very short intervals where the local fluctuations are dominant. We indeed found non uniform characteristics of the segments. The variability (SD) of CDS length series for *Hahella chejunesis* and *Bacillus thuringiensis* looks quite non uniform among the ten segments (figure 7-8). In the first case the highest variability appears to be concentrated in the middle of the CDS length series while in the second case at the margins. The correlation exponent $\alpha_1$ and $\alpha_2$ also appear to be highest in the middle of CDS length series for *Hahella chejunesis* so they parallel the variability of the distribution. To a first approximation correlation is stronger where variability is greater. Therefore statistical properties (distribution and correlation) of the whole CDS length series represent averaged values, consequently they give a limited and quite general information about their organization.

Figure 5 Detrended fluctuation analysis of the coding sequences length series of *Clostridium beijerinckii.* The lower plot is for the same series subjected to shuffling. The correlation exponents are included in the figure. The shuffled data are described by a single correlation time α.

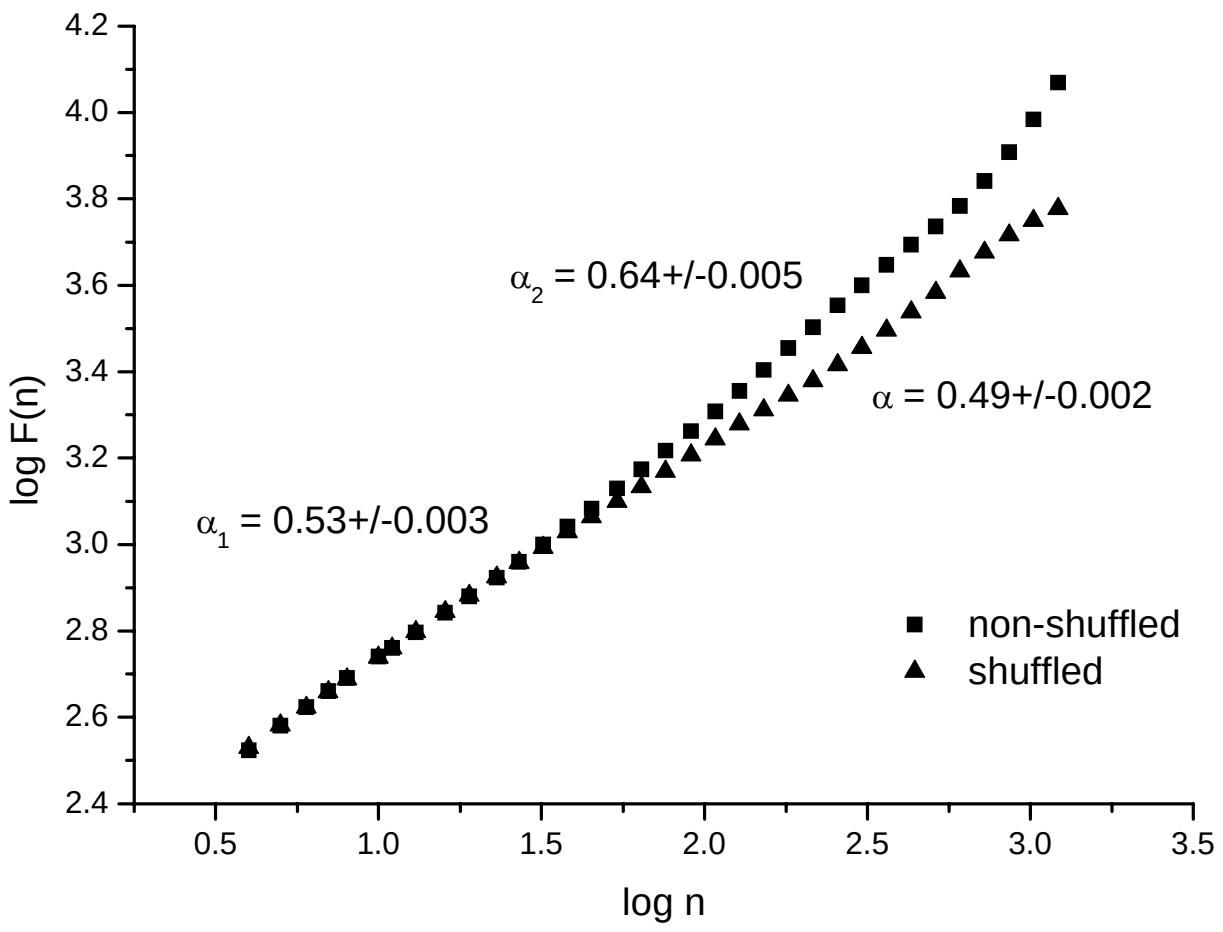

Figure 6. The short distance $\alpha_1$ and long distance $\alpha_2$ correlation exponents for microbial CDS length series. Data for two natural languages and a computer language are also included. The straight line is intended to guide the eye in relationship to the data for natural and computer languages. Some of the species are also identified.

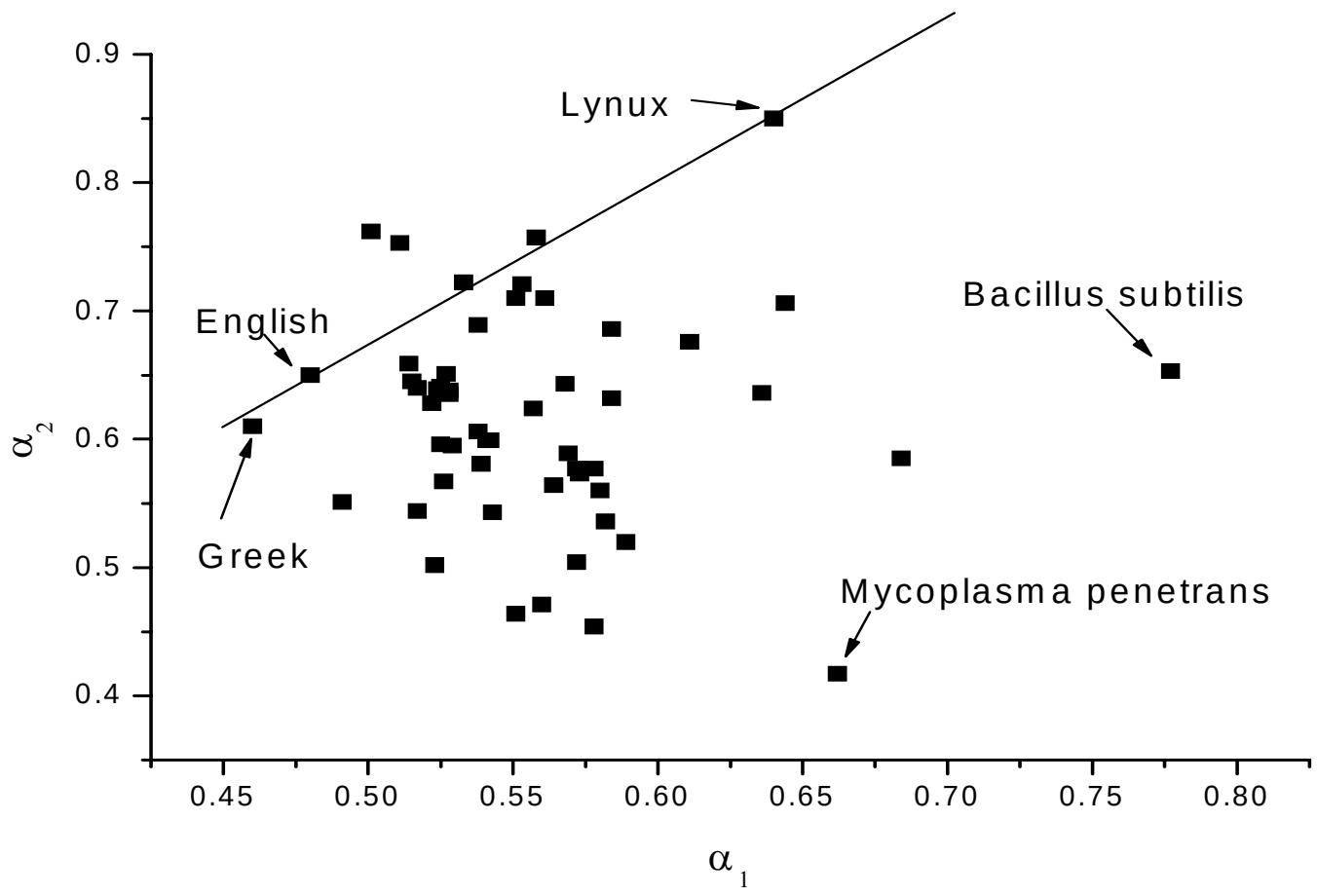

Figure 7 Variability (SD) and correlation exponents ($\alpha_1$ and $\alpha_2$) for ten segments of CDS length series of *Hahella chejuensis* bacteria.

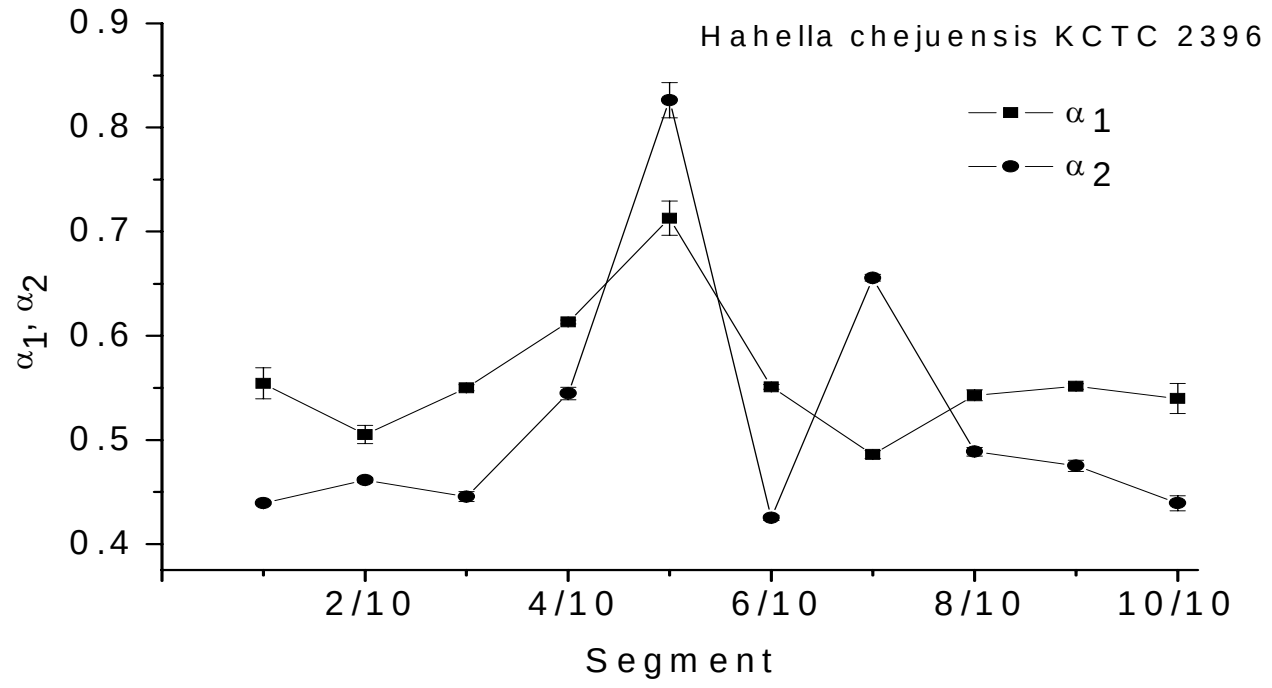

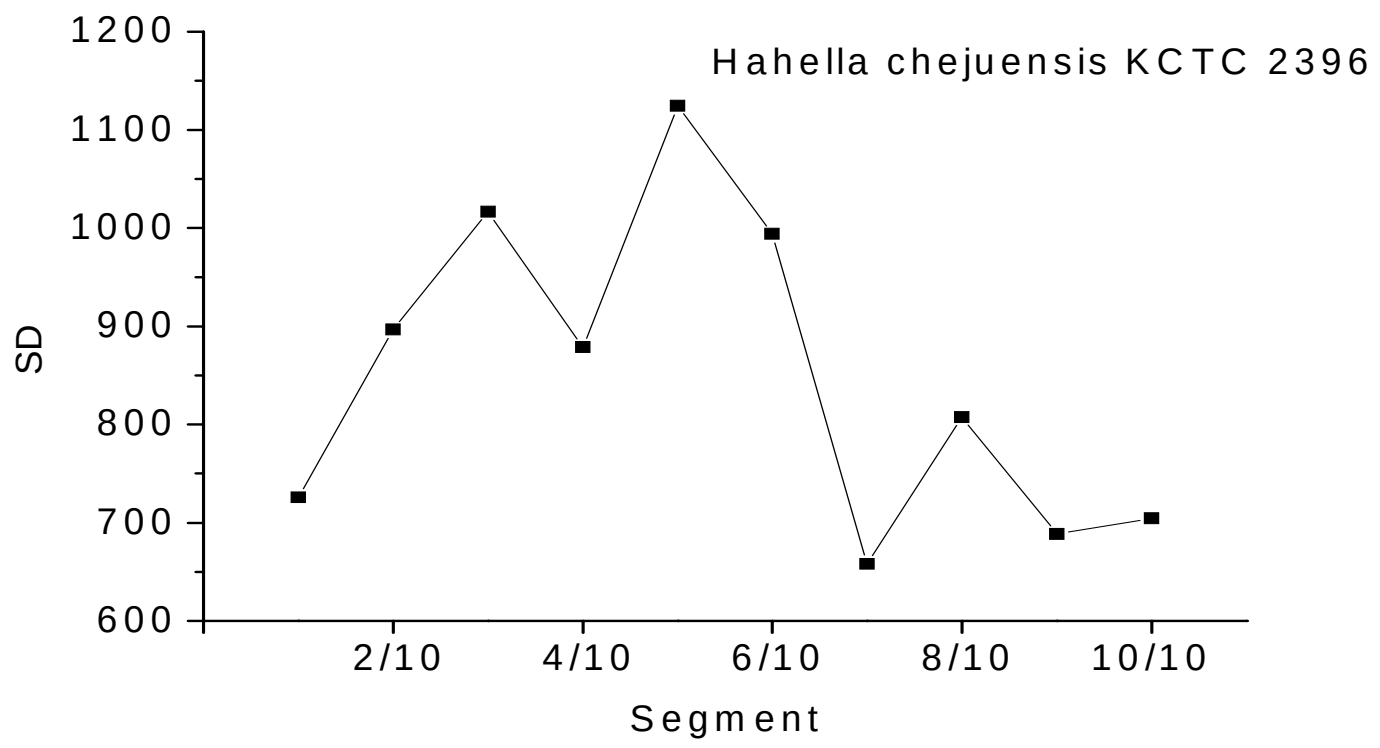

Figure 8 Variability (SD) and correlation exponents ($\alpha_1$ and $\alpha_2$) for ten segments of CDS length series of *Bacillus thuringiensis* bacteria.

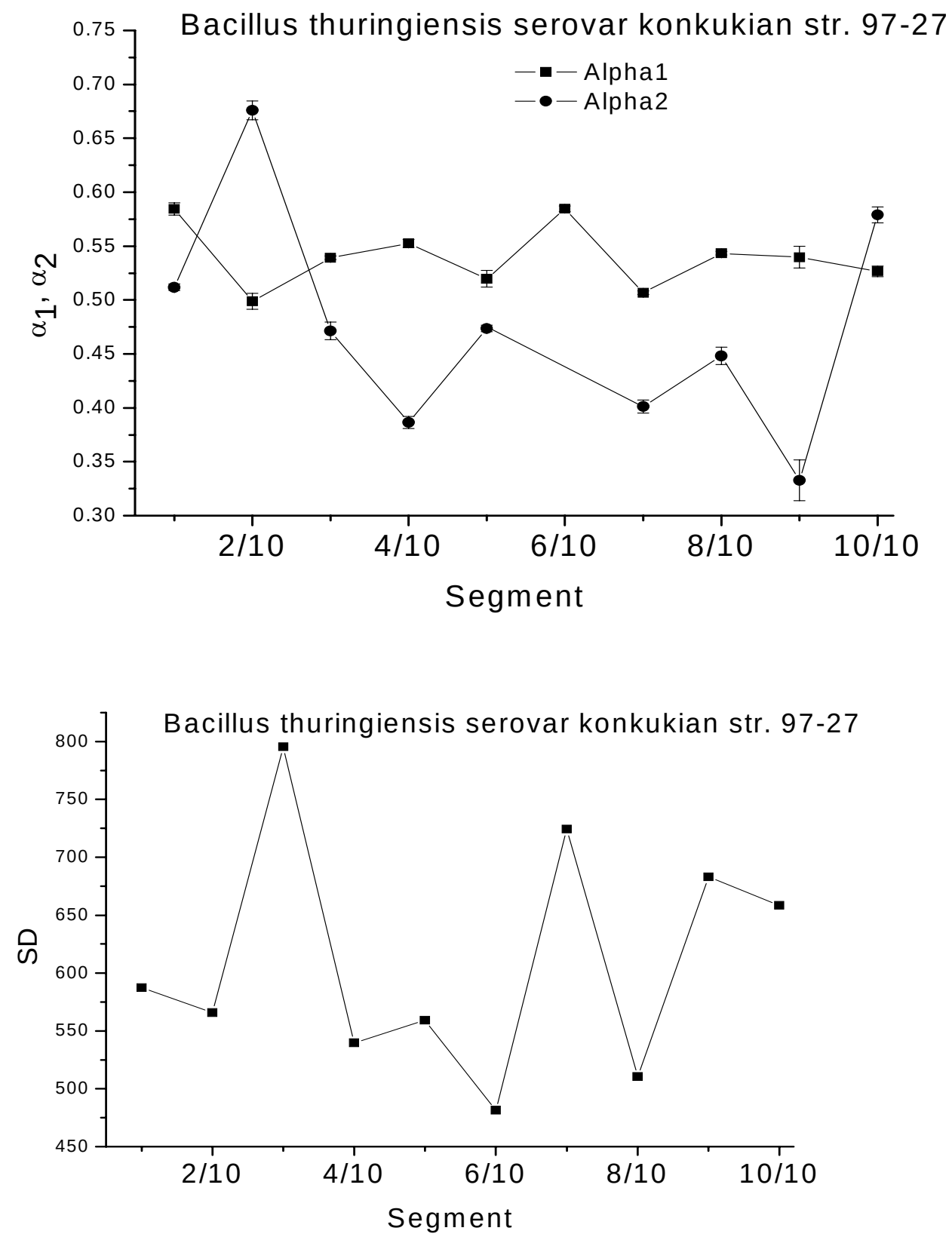

In terms of correlation a non uniform organization of the series is equivalent to short-range correlation. This is understood as a correlation $\alpha \neq 0.5$ extending over less than about one order of magnitude of box sizes *n* [8]. In other words CDS length values appear to be slightly or stronger correlated at short distances or at longer distances. Here "long" distances should not be confounded with "long range" correlation. We remind again that "long range" correlation means correlation over any scale in the series (also known as fractal) while correlation at "longer distances" remains within the short-range correlation class. To be more explicitly short-range correlation at "short distances" means for example that correlation between the first and second, first and third,…, first and 9th term are higher than 0.5. Also short-range correlation may hold at longer distances, for example between the first term and the $10^{th}$ term, between the first and the $11^{th}$ term, the first and the $12^{th}$ term and so on. This example is clearly a short-range correlation as it extends over a limited range of data, lower than an order of magnitude. In short-range models like autoregressive processes the correlation or memory of the first term decreases as distance increases [8]. In fact a further line of investigation of the non-uniform organization of the microbial genome is to apply autoregressive models. The simplest model is the first degree autoregressive process (AR1) given by the equation:

$$X_t = \varphi X_{t-1} + \varepsilon_t \qquad (4)$$

where $\varepsilon_t$ is a white noise process with zero mean and variance $\sigma^2$, while φ is a parameter. The parameter vales φ have to be restricted for the process to be stationary which means that $|\varphi|<1$. If $\varphi=1$ then $X_t$ can also be considered as a random walk. Parameter φ can be regarded as the strength of interaction among the terms $X_i$ [8]. Obviously the more distant the terms of the series the lower is the correlation. Regardless of a temporal or spatial process the parameter φ can be understood as the scale of short range memory of the system. Higher order models AR(p) are characterized by more *p*arameters of $\varphi_i$ which indicate the strength of interaction between successive nearby terms and more distant terms. AR models have successfully been applied to astrophysical and psychological data [13-14]. More recently we found that various biophysical phenomena can be well described by AR(1) models or by higher order AR(p) models [unpublished work]. They include the structure of proteins, flickering of the red blood cells and random number generation by human subjects. It was however felt that a systematic DFA of a short range memory model is needed to better understand how the correlation and the scale of memory are related.

## 6. Conclusions

The overall statistical characterization of the CDS length series show that distribution is exponential or approximated by exponential like distribution. This is unlike the distribution of words in natural languages which obeys Zipf's law.

The correlation of data in the series is weak at distances shorter than an order of magnitude and more pronounced (either stronger or anti correlated) at longer distances than an order of magnitude. The correlation of words for natural and computer languages fall into the same area of values. This shows that the language of microbial genome and that of natural languages have both significant difference and resemblance at the same time. While the overall characteristics for the whole genomes offer a broad idea of the statistical properties of the genome it is felt that further insight is necessary to understand their organization. The reason is that the CDS length series seems to have a non uniform structure. It looks like that distribution and correlation of the data change along the genome and therefore cannot be fully characterized by an overall distribution and correlation characteristics like in a long range correlated system. Segmentation of the genome into several intervals revealed that distribution and correlation may significantly rise at certain locations along the genome. This depends, in turn, on the species. It is suggested that the non-uniform organization of the coding sequences in the microbial genomes could be modeled by autoregressive processes which are typical short-range memory systems.

## Acknowledgements

This work was supported by a grant from the National Authority for Scientific Research.

## References

[1] Yu, Z.-G., Anh, V.: Time series model based on global structure of complete genome, Chaos Solitons and Fractals. **12**, 1827-1834 (2001)

[2] Yu, Z.-G., V.Anh, Wang, B., Correlation property of length sequences based on global structure Z.-G. of the complete genome. Phys. Rev.E. **63,** 011903 (2001)

[3] Zainea, O., Morariu, V.V., The length of coding sequences in a bacterial genome: Evidence for short-range correlation. Fluct.Noise Lett. **7,** L501-L506 (2007)

[4] Zainea, O., Morariu, V.V., Correlation investigation of bacterial DNA coding sequences. Romanian J. Biophys. **18** (2008) in print

[5] The University of Tokyo, Centre of Excelence Programs, Elucidation of language structure and semantics behind genome and the life system, http://www.cb.k.u-tokyo.ac.jp/coe/index-en.html, cited 20 March 2008 (2008)

[6] Brakman, S., Garretsen, H., Van Marrewijk, C., Van Den Berg, M., The Return of Zipf: Towards a Further Understanding of the Rank-Size Distribution. Journal of Regional Science. **39:** 183-213. DOI 10.1111/1467-9787.00129 (1999)

[7] Peng, C.-K., S.V. Buldyrev, S.V., Havlin, S., Simons, M., H.E. Stanley, H.E., Goldberger, A.L., Mosaic organization of DNA nucleotides. Phys. Rev. E. **49,** 1685-1689 (1994)

[8] Morariu, V.V., Buimaga-Iarinca, L., Vamoş, C., Şoltuz, Ş.M., Detrended fluctuation analysis of autoregressive processes. Fluct.Noise Lett.**7,** L249-L255 (2007)

[9] Ferrer i Cancho R., Sole, R.V., Zipf's law and random texts. Advances in Complex Systems. **5**, 1-6 (2002)

[10] Steels, L., Analogies between Genome and Language Evolution. In: Pollack, J. et.al. (eds) Proceedings of Alife 9. The MIT Press Cambridge Ma (2004)

[11] Morariu, V.V., A limiting rule for the variability of coding sequences length in microbial genomes, *arXiv*: 0805.1289 v [physics.bio-ph ]

[12] Kosmidis, K., Kalapokis, A., Argyrakis, P., Language time series analysis, Physica A. **370** 808-816 (2006)

[13] König, M., Timmer, J., Analyzing X-ray variability by linear state space models. Astron. Astrophys. Suppl. Ser. **124**, 589-596 (1997)

[14] Thornton, Th. L., Gilden, D., Provenance of correlation in psychological data. Psychonomic Bulletin & Rev. **12,** 409-441 (2005)